\documentclass{PoS}
\usepackage{subfigure,epsfig}

\title{Thermodynamical properties of QED in 1+1 dimensions within light front dynamics}

\ShortTitle{Thermodynamical properties of QED in 1+1 dimensions within light front dynamics}

\author{\speaker{Stefan STRAUSS}%
         \thanks{We thank the conference organizers of Light Cone 2008 for the opportunity to present our work and the inspiring atmosphere during the meeting.}\\
        University of Rostock, Germany\\
        E-mail: \email{stefan.strauss@uni-rostock.de}}

\author{Michael Beyer\\
        University of Rostock, Germany\\
        E-mail: \email{michael.beyer@uni-rostock.de}}

\abstract{We investigate thermodynamical properties of quantum
electrodynamics in $1+1$ dimensions (QED$_{1+1}$). Discrete light cone
quantization is used to compute the partition function of the
canonical ensemble and the thermodynamical potential. The potential is
evaluated for different system sizes and coupling strengths. We
perform the continuum limit and the thermodynamical limit and
present basic thermodynamical quantities as a function of temperature
for the interacting system. A more accurate estimation of low lying
bound state masses at non-perturbative coupling strength are determined
due to the higher harmonic resolution. The results are
compared to the idealized cases.}

\FullConference{LIGHT CONE 2008 Relativistic Nuclear and Particle Physics\\
                 July 7-11, 2008\\
                 Mulhouse, France}

\begin{document}

\section{Introduction}
Recently thermal field theory in the light front (LF) frame introduced by Dirac \cite{Dirac:1949cp} has gained quite some attention. The most important application in this 
framework is the phase diagram of strongly coupled systems like, e.g., the
quark gluon plasma. Todays perception of the QCD phase diagram is
due to by lattice QCD computations. However, these calculations are limited to the region
$T\le\mu$ ($T$ temperature, $\mu$ quark chemical potential) due to the
complex action at large chemical potential. In turn, this results in the well known sign problem of the Monte-Carlo simulation method. The generic Monte-Carlo sign problem is at least as hard to compute as problems in the complexity class NP (class of non-deterministic polynomial problems) and every problem in NP is reducible to the sign problem in polynomial time (i.e. the generic Monte-Carlo sign problem is NP-hard)~\cite{troyer-2005-94} 
and therefore we argue that it is worth looking for alternative ways to determine the QCD phase diagram.

In the following we investigate LF quantization to compute thermodynamical quantities. The first attempt to use results of light front quantization, i.e. the
invariant mass spectrum and the wave functions of the theory, for
applications in thermodynamics has been given in
\cite{Elser:1996tq}. However, the conclusions of Ref.~\cite{Elser:1996tq} are rather confusing since a second order phase
transition in one-dimensional QED has been conjectured. The limiting cases
of non-interacting fermions on one hand and the free boson gas on the other
have not been considered. Certain
classes of supersymmetric models \cite{Hiller:2007sc}
and four-dimensional pure gluonic QCD \cite{Dalley:2004ca} have been
investigated and thermodynamical properties computed. 
Analytical calculations in LF thermal field theory have been performed for different
models. These perturbative computations have been done using a
statistical operator familiar from the more traditional instant form approach. It
was possible to reproduce known results like thermal masses in scalar
field theory and properties of the Nambu-Jona-Lasino model
\cite{Alves:2002tx,Beyer:2001bc}. A notation of the general light cone 
(GLC) frame, which compromises between instant and front form coordinates, was
introduced \cite{Weldon:2003uz} and in the following it was pointed
out that the canonical quantization in the GLC frame \cite{Das:2004je}
is essentially analogous to ordinary light cone quantization. However, the
advantages of light cone quantized thermal field theory stemming from
technical simplifications in perturbative computations like the simple
pole structure of the propagator have been hardly exploited, see e.g. \cite{Weldon:2003vh}.

A non-perturbative approach to light cone quantized field theories is
given by discrete light cone quantization (DLCQ)
\cite{Eller:1986nt}. Discrete light cone quantization is a finite box quantization
of Hamiltonian field theory supplemented by boundary conditions for
the fields and cuts the Fock space into finite-dimensional sectors of
equal resolution $K=\frac{L}{2\pi}P^+$, where $L$ is the box
length. Mass spectra and LF wave functions of low lying states which
are independent of the box length have been numerically computed for
one-dimensional or dimensionally reduced systems via DLCQ. Higher
dimensional systems are usually treated by the transverse lattice approach which
replaces two spatial dimensions by a lattice and the remaining two by
DLCQ. The problem of renormalization in Hamiltonian field theory and
therefore the construction of effective light cone Hamiltonians
remains to be solved and hampers application of light cone
quantization to non-perturbative quantum field theory in
3+1 dimensions.

In this proceeding we carefully reconsider questions in QED$_{1+1}$ as raised in
\cite{Elser:1996tq}. However arrive at mostly different conclusions concerning the conjectured 
phase transition. Some of our results have been recently given in Ref.~\cite{Strauss:2008zx}.

\section{Light Front Thermodynamics}
Starting from considerations in Ref.~\cite{Raufeisen:2004dg} 
the statistical operator on the light front can be written in the following form 
\begin{equation}
\label{eq:statisticaloperator}
\varrho=\frac{1}{\cal Z} \exp\left\{\frac{\beta}{2}\left(P^+ + P^-\right)\right\},
\end{equation}
apparently different from the 'naive' light cone version $\varrho_{LC}\sim\exp(\beta P^-)$ that resembles the non-relativistic form.
The partition function is given by ${\cal Z}={\rm Tr}\varrho$ which is the central quantity when one wants to compute thermodynamical and
statistical properties. When evaluating the partition function in DLCQ one introduces the harmonic resolution $K$ and the light cone Hamiltonian $H$ through
\begin{equation}
\label{eq:p^+,-}
P^+ = \frac{2\pi}{L}K,\qquad P^- = \frac{L}{2\pi}H=\frac{L}{2\pi}\frac{M^2}{K}.
\end{equation}
Here $K$ is dimensionless, diagonal in the DLCQ basis and used as a
measure of the discrete approximation. The light cone Hamiltonian has
dimension mass squared and is the dynamical part in
(\ref{eq:statisticaloperator}) since it is a non-diagonal matrix of
increasing size in $K$. Inserting (\ref{eq:p^+,-}) the partition
function reads
\begin{equation}
\label{eq:partitionfunction}
{\cal Z}(T,L) = {\rm Tr}\; \exp\left\{
\frac{\beta}{2}\left(\frac{2\pi}{L} K*\hat{I} + \frac{L}{2\pi} M_0^2
\frac{\hat{M}_K^2}{K}\right)\right\},
\end{equation}
where '${\rm Tr}$' means summing over all resolutions $K$ and all
corresponding (decoupled) Fock space sectors, and $\hat{I}$ is the
identity matrix. The mass matrix $\hat{M}_K$ of course is different
for different \textit{K}-sectors. Note that the volume appears explicitly in (\ref{eq:partitionfunction}) in contrast to the suggestion for ${\cal Z}$ in \cite{Elser:1996tq}. This is due to the consistent approach based on eq. (\ref{eq:statisticaloperator}).
$M_0$ is the mass of the lightest state in
the continuum limit, that means we normalize the smallest eigenvalue
of $\hat{M}_K$ to one for $K\rightarrow\infty$. In a numerical
computation we fix the volume (in units of the continuum estimate of
the lowest mass) at the beginning and extrapolate for $K$ to
infinity. This calculation has to be performed for several values of
$L$ to safely determine the expected linear dependence
\begin{equation}
\label{eq:linearscaling}
\Omega=-T\ln{\cal Z}= \alpha L + \beta,
\end{equation}
where $\Omega$ is the thermodynamical potential. We
emphasize that one has to pick a strict order of limits in $L$ and
$K$, first take $K\rightarrow\infty$ followed by
$L\rightarrow\infty$. Computing (\ref{eq:partitionfunction}) in practice means exponentiating large matrices and
summing the diagonal elements. For small resolutions $K$ this is most
conveniently done by first computing the eigenvalues and then the
matrix exponential. At larger resolutions we employ a random vector
routine \cite{Iitaka} to compute the trace of the matrix exponential,
which has been approximated by Trotter decomposition.

\begin{figure}[t]
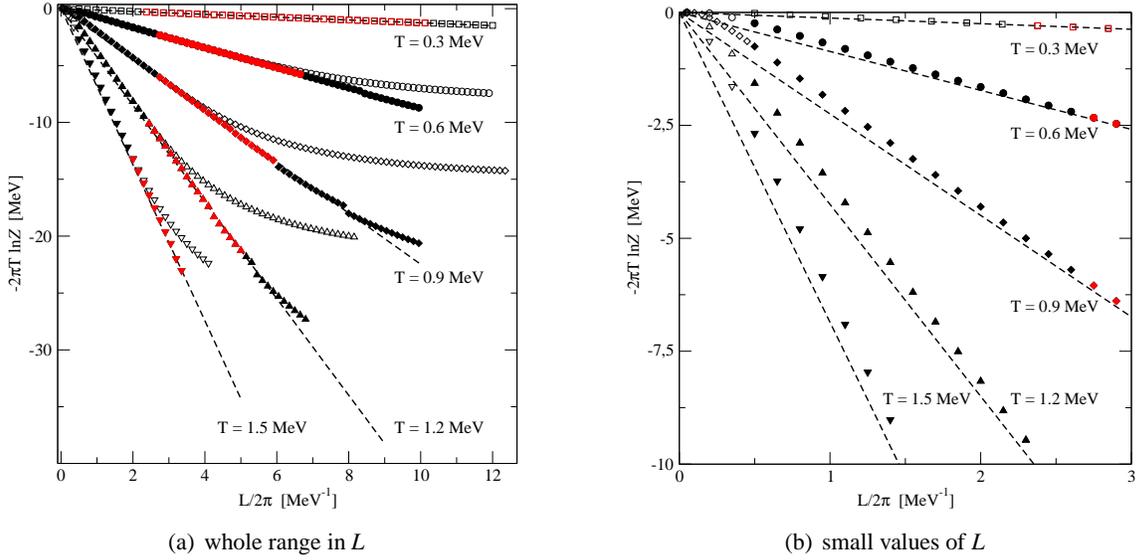

\subfigure[whole range in $L$]{\includegraphics[width=0.45\textwidth]{freethpot.eps}
\label{fig:freethpot}}\hfill
\subfigure[small values of $L$]{\includegraphics[width=0.45\textwidth]{freethpot_insert.eps}
\label{fig:finitesize}}
\caption{
The thermodynamical potential density $-2\pi T\ln\mathcal{Z}$ as a
function of $L$ for different temperatures. In figure (a) the whole
interval in $L$ is shown while in figure (b) $L$ is limited to small
values to present the finite size effects. Open symbols in both
figures are the potential at maximal resolution $K=110$. Closed
symbols are given by an extrapolation. The slope of the linear part
(values selected colored in red) is fitted to extract the invariant
potential density.}
\end{figure}

As a test case and to fix the range of external parameters $T, L$
where reliable numerical results can be
extracted we investigate the free Fermi gas. The light cone expression
of the thermodynamical potential (density) of the free quantum gases
is given as (upper sign fermion ($f$), lower sign boson ($b$))
\begin{equation}
\label{eq:lc_freepot}
 \omega_{f/b} = \mp T\int\limits_0^\infty
\frac{dp^+}{2\pi}\ln\left(1\pm\exp\left\{-\beta\left(\frac{p^+}{2}+
\frac{m^2}{2p^+}\right)\right\}\right).
\end{equation}
Equation (\ref{eq:lc_freepot}) is derived analogous to the instant
form case, replacing the spatial volume by the light-like
extension. In the large 'volume' limit the densities are equal in both
relativistic forms. Figure \ref{fig:freethpot} shows results for the
free electron gas of mass $m=0.5$ eV at resolution $K=110$. At small
system volumes clear finite size effects are visible (see figure
\ref{fig:finitesize}) and at large volumes there are derivations from
the exact result (\ref{eq:lc_freepot}) because of the finite
resolution. Therefore one has to identify a scaling window where the
linear behavior in (\ref{eq:linearscaling}) shows up. Finding such a
window is easy at small temperatures, but for increasing temperatures
the scaling window is pushed to regions of large volumes. For the largest temperature shown in figure \ref{fig:freethpot} the relative error is below $1.5 \%$, see \cite{Strauss:2008zx} for more details.

\section{QED$_{1+1}$ at finite Temperature on the Light Front}
\begin{figure}
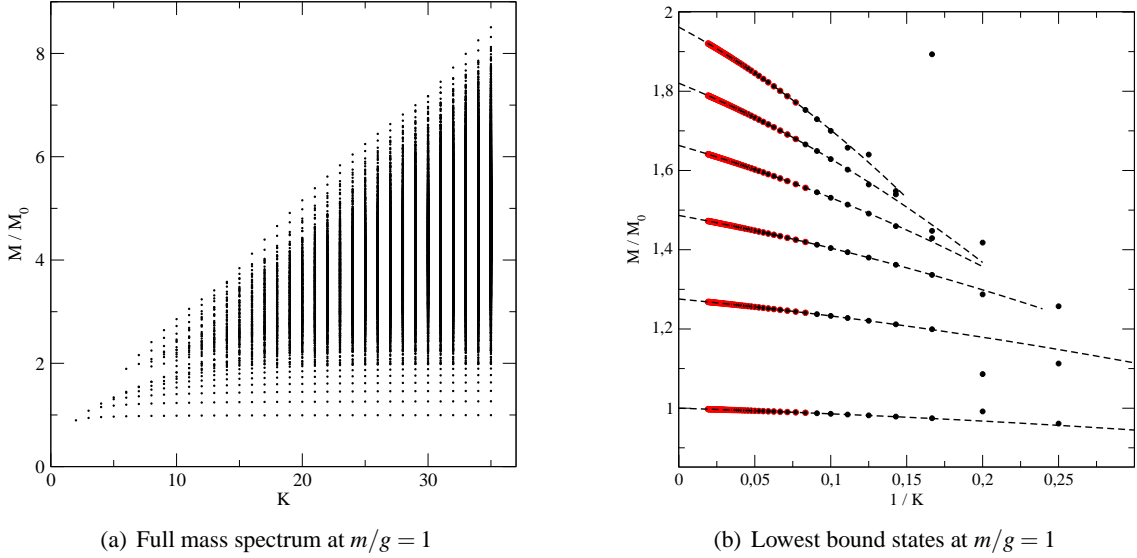

\subfigure[Full mass spectrum at $m/g=1$]{\includegraphics[width=0.45\textwidth]{new_spectrum.eps}
\label{fig:mass(A)}}\hfill
\subfigure[Lowest bound states at $m/g=1$]{\includegraphics[width=0.45\textwidth]{sixmasses_g0_trans.eps}
\label{fig:mass(B)}}
\caption{The invariant mass spectrum at $m/g=1$.
Part (a) shows the full mass spectrum up to $K=35$ and in (b) the six
lowest mass eigenvalues of QED$_{1+1}$ are depicted. The dashed line
is a quadratic fit to the data (values used in the fit are colored in
red) and used to extract the continuum limit. $M_0$ is defined such that
the continuum value of the lowest mass is normalized to one. }
\label{fig:mass}
\end{figure}

The light front Hamiltonian of the massive, chiral Schwinger model
(QED$_{1+1}$) is given in \cite{Eller:1986nt} without the dynamical
gauge field zero mode. Generically the Hamilton operator has the structure
\begin{equation}
H=m^2H_0+\frac{g^2}{\pi}V=g^2\left(\frac{m^2}{g^2}H_0+\frac{1}{\pi}V\right),
\end{equation}
where $H_0$ is the free Hamiltonian, that is diagonal in free particle basis and $V$ some complicated operator containing combinations of
four creation and destruction operator of fermions and
anti-fermions. The application of DLCQ to thermodynamics requires
rather larger harmonic resolutions, as a byproduct one gets more
accurate estimates for mass spectrum for different couplings. Still
one has the extrapolate the raw data to the limit
$K\rightarrow\infty$, which was done by second-order power functions
in $1/K$. In figure \ref{fig:mass} the mass spectrum is plotted for
$m/g=1$, which is in the non-perturbative coupling regime. Thereby
\ref{fig:mass(A)} shows the full spectrum up to $K=35$ and the growth
of DLCQ states is apparent. For $M/M_0> 2$ the spectrum is continuous and we singled out 
the six lowest mass states in figure \ref{fig:mass(B)}.

A comparison with masses obtained by other means like finite
lattice calculations \cite{Sriganesh:1999ws},
variational DLCQ \cite{Mo:1992sv} and fast moving frame approach
\cite{Kroger:1998se} is possible for the lowest two states and our
results \cite{Strauss:2008zx} are generally in very good agreement.
Slight differences appear for $m/g\le 2^{-3}$. The reason is that the choice of the fermionic Fock representation is presently not optimal in this case.

\begin{figure}
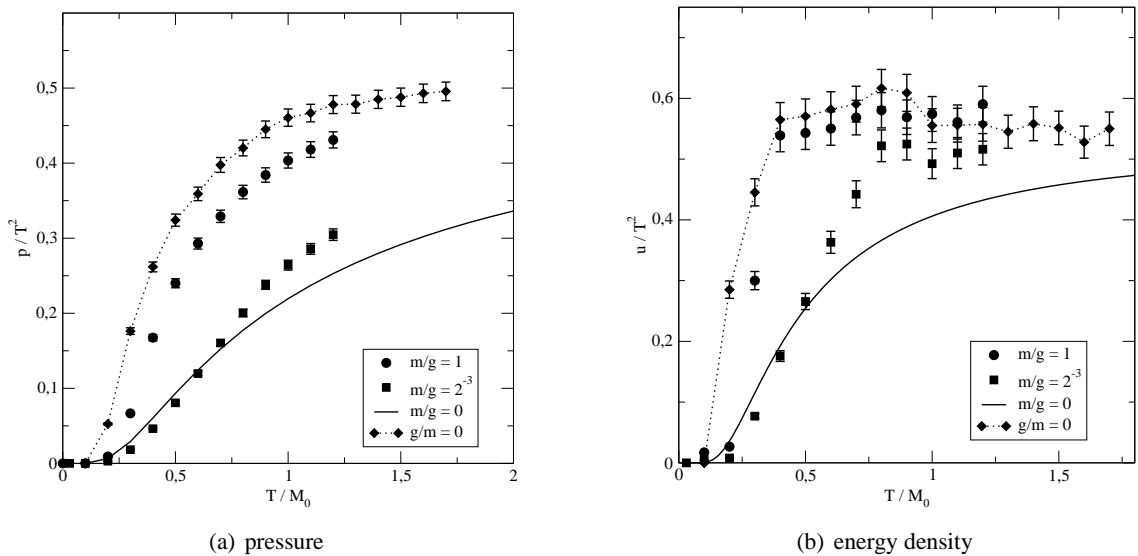

\subfigure[pressure]{\includegraphics[width=0.45\textwidth]{Druck_chg.eps}
\label{fig:Druck}}\hfill
\subfigure[energy density]{\includegraphics[width=0.45\textwidth]{energy_chg.eps}
\label{fig:energy}}
\caption{The thermodynamical quantities pressure (a) and internal energy density 
(b) divided by $T^2$ as functions of temperature. Four different
couplings are shown: pure Bose gas (solid line), strongly interacting
Fermi system $m/g=2^{-3}$ (squares), weaker interacting Fermi system
$m/g=1$ (circles), free Fermi system $g/m=0$ (diamonds).}
\label{fig:freepotfull}
\end{figure}

The thermodynamic quantities are obtained in the way outlined in the former section.
Following two relations hold 
\begin{equation}
p=-\omega=\frac{T}{L}\ln{\cal Z}\quad\mbox{and}\quad
u=\frac{\partial\ln{\cal Z}}{\partial T} \frac{T^2}{L}.
\end{equation}
In practice, we have directly computed the pressure and used a numerical derivative to determine the internal energy density
$u$. Figure \ref {fig:Druck} (\ref{fig:energy}) show the dimensionless ratio of pressure (internal energy density) and $T^2$ as a function of
temperature of a QED gas for four different couplings. For massive
fermions we meet the chargeless condition of physical states
$Q|phy\rangle=0$ and thus computed ${\cal Z}$ in the canonical
ensemble. In the limit of vanishing mass we used grand canonical
ensemble with $\mu=0$ since in this case LF QED$_{1+1}$ is a free boson theory
of mass $m_b=\frac{g}{\sqrt{\pi}}$. The errors in figure
\ref{fig:Druck} are due to the extrapolation to larger resolutions and 
roughly carried over from the free case computation in section 2. In figure \ref{fig:energy}
the fluctuations of the data points can be reduced by setting a
smaller temperature grid. The external parameters $T, L$ are both
given in units of the lowest bound state mass $M_0$. Unlike to the free case
before we do not set a definite physical scale since $M_0$ is not
fixed in physical units. Remind that the mass of the first bound state
can be large, like in QCD where the lightest bound state is the pion of
mass $m_\pi=140$ MeV made out of nearly massless quarks. To judge whether
the temperature reached in the numerical computation is sufficient we
compare the high-temperature values in the figures
\ref{fig:freepotfull} with the $T\gg 0$ regime of
(\ref{eq:lc_freepot}).  One finds that the pressure has not yet
reached the value expected by the high-temperature limit $p/T^2\approx
\pi/6$, but the internal energy is at $T\ge M_0$ in the range of 
$u/T^2\approx\pi/6$. In comparison to the results of the earlier study
\cite{Elser:1996tq} $p$ and $u$ are computed at significantly higher
temperatures and no sign of the conjectured phase transition is
found.  The figures \ref{fig:freepotfull} offer the
interpretation that the thermodynamical quantities change smoothly
under variation of the coupling. 

\section{Conclusion}
This contribution is concerned with the application of light cone
quantization to the thermodynamics of non-perturbative quantum field
theory. As an example we treated QED in $1+1$ dimensions and presented
the pressure and the internal energy. Since we have computed the
partition function other thermodynamical quantities like entropy and
the specific heat can be obtained via derivatives of $\ln{\cal Z}$ and
the equation of state can be given numerically. This procedure is
limited by the exponential growth of basis states and the
dimensionality of the Hamiltonian matrix with increasing harmonic
resolution.  To this end an effective (non-perturbative)
renormalization program for Hamiltonians is necessary. Promising
suggestions to this direction are the similarity transformation
renormalization transformation
\cite{Glazek:1993rc,Gubankova:1997mq}, and the density matrix renormalization group in momentum space
\cite{MartinDelgado:1999jb}. More specifically within the massive Schwinger model an inclusion 
of the dynamical zero mode is desirable because the condensate
connected to the zero mode may have impact on the thermodynamics. In
Ref. \cite{Martinovic:1998pr} such a light cone Hamiltonian is suggested which could be a good starting point.

A main objective of this direction of research is the extension to
four-dimensional finite density QCD, avoiding the Monte-Carlo sign
problem and reveal how the phase diagram of quark matter is from first principles. So far we have proven that the consistent application of the  theoretical framework outlined in \cite{Raufeisen:2004dg} to the non-perturbative situation is possible and leads to reasonable thermodynamic results. The numerical issues faced will surely increase, if one considers the full 3+1 case. Nevertheless, the chance of eventually arriving at results for the phase diagram of QCD alternative to the ones of the well established lattice QCD is really exciting and along the demands recently claimed by Ken Wilson~\cite{Wilson:2004de}.

\end{document}